# Gen-AI-tecture: using generative AI to support architectural students in design tasks

## Authors

Timo Kapsalis [ORCID: 0000-0003-1598-6426][(1)]

(1): College of Science & Engineering, University of Derby, UK

## Abstract

The *Gen-AI-tecture* project embeds a locally executed, discipline-specific tool into a mixed-methods focus-group design, structured around three research objectives: (RO1) to evaluate how generative AI tools impact students' creativity in design-thinking processes and outcomes, (RO2) to assess whether these tools enhance inclusivity in learning processes, and (RO3) to examine how they develop students' AI-handling skills with a view to boosting future employability. Findings indicate enhanced creative fluency, broadened participation across diverse learner profiles, and strengthened confidence in AI-supported design processes. The study contributes evidence-based guidance for integrating generative-AI workflows into architectural pedagogy, demonstrating how such tools can operationalise constructivist principles of learner-led meaning-making, support connectivist understandings of learning as participation in human-AI networks, and advance universal learning theories by promoting more inclusive, flexible and accessible educational practices for contemporary learners.

# 1. Introduction

The rapid advancement of generative artificial intelligence (GenAI) offers transformative potential for architectural education, yet empirical studies on bespoke, discipline-specific tools remain limited. Much existing scholarship concentrates on generic platforms, leaving a gap around custom AI systems aligned with architectural workflows and studio cultures. By developing and evaluating tailored GenAI tools, recent work addresses four recurring needs: robust technical development, demonstrable effects on students' creativity, contribution to inclusive learning environments, and cultivation of AI-handling skills linked to employability (Aly et al., 2025; Jin et al., 2024; Sadek & Mohamed, 2023).

Generic AI platforms enhance accessibility but often lack architectural specificity and contextual awareness (Aly et al., 2025). Bespoke systems trained on architectural datasets and embedded within design workflows can support automated design generation, targeted feedback and basic code-compliance checking (Aly et al., 2025; Jin et al., 2024). Alamasi et al. (2026) offered controlled evidence from a supervised studio project in which GenAI integration improved performance relative to a control group, particularly in design form and aesthetics, with a 14% gain and associated improvements in time efficiency and design quality. Their framework proposes graded levels of GenAI use across design stages to preserve critical engagement.

Empirical studies suggest that generative AI can augment creativity and task efficiency when carefully integrated into studio pedagogy. Sadek & Mohamed (2023) showed that students using AI-driven image-generation tools produced more innovative conceptual designs than those relying on traditional methods, while Jin et al. (2024) reported improved problem-solving and conceptual understanding in an AI-assisted course. Choo, Park & Hong's (2025) two-year analysis of courses using Stable Diffusion demonstrated that, as students combined generative tools with conventional software like Photoshop, Rhino, Revit and SketchUp, AI shifted from a peripheral visualiser to a central part of the creative process, strengthening understanding of architectural workflows. Nonetheless, these studies highlight persistent challenges, including steep learning curves, fragmented toolchains and unpredictable outputs, which point to the need for scaffolded training and explicit studio protocols.

AI's adaptive capabilities can personalise feedback and accommodate diverse learning preferences, potentially reducing barriers for novices and students with different learning profiles (Melo-López et al., 2025). Generative tools help students with weaker drafting or visualisation skills articulate complex spatial ideas, democratising participation (Aly et al., 2025; Paananen, Oppenlaender & Visuri, 2023). Huh, Miri & Tracy's (2025) exploratory study of architecture and interior architecture students reported strong enthusiasm for generative image tools in early-stage design, where they were perceived as boosting creative confidence and professional competitiveness. Yet acceptance declined in later stages, where originality and critical judgement were seen as paramount, and participants stressed the importance of transparent acknowledgement of AI use. These findings resonate with concerns that unequal access to high-performance hardware and algorithmic bias may exacerbate rather than alleviate inequities if not proactively managed (Melo-López et al., 2025).

Finally, the architecture, engineering and construction (AEC) sector's growing reliance on AI foregrounds the importance of cultivating "AI-handling" competencies within architectural curricula. Jin et al. (2024) showed that structured exposure to AI workflows develops conceptual and practical skills, while survey evidence indicates that nearly 70% of architecture students already use AI tools independently, despite more than 95% reporting no formal AI education (Dullinja & Jashanica, 2025). The recent "AI Report" from the Royal Institute of British Architects (RIBA, 2025) notes that over half of practices now integrate AI into projects. Taken together, these studies suggest that embedding AI literacy into design education is increasingly essential for employability, provided that pedagogic models also foreground ethics, inclusivity and critical reflection.

It is within this landscape that the *Gen-AI-tecture* project situates itself, examining a bespoke, locally run generative-AI workflow designed to support creativity, inclusivity and AI-handling skills in undergraduate architectural education. The developed tool enables students to rapidly generate and revise visualisations of interior spaces by describing their ideas in natural language and making targeted adjustments to existing images. Rather than replacing design thinking, the system acts as a visual co-pilot. That is, it helps students explore multiple alternatives, compare options, and refine aesthetic and spatial decisions. In doing so, the developed tool lowers technical barriers to high-quality visual output while foregrounding reflection, iteration and critical judgement.

# 2. Methodology

## 2.1. Description of the developed tool and workflows

The Gen-AI-tecture intervention centres on a bespoke generative-AI image creation and editing workflow designed specifically for architectural education. Technically, the system is implemented as a localised *ComfyUI* workflow. ComfyUI (ComfyUI, 2025) is a node-based software application, in which AI models and operations are connected visually. From a user's point of view, it functions as a relatively simple web-like environment in which students upload or select images, describe desired changes in natural language, and receive multiple visual alternatives in response.

The workflow is powered by the Flux 1 Kontext (dev) diffusion model (Black Forest Labs, 2025), which has been fine-tuned with a curated set of interior-design images. In this context, fine-tuning refers to the process of "teaching" a general AI model a specialty (Fullan et al., 2018) - i.e., in this case, urban design imagery - using an add-on method called *Low Rank Adaptation (LoRA)*. LoRA works like a lightweight plug-in, as it adapts the base model to the specialty domain without retraining it from scratch (Luo et al. 2025). For the fine-tuning process, we used the *InteriorNet* dataset (Li et al., 2018), which includes 20 million interior shots so that our system is more adept at rendering room-scale spatial qualities, materials, lighting conditions and furniture arrangements typical of residential interiors. As a result, the system can produce coherent bedroom and living/dining room scenes with sufficient fidelity to support early design discussions without claiming the status of resolved architectural drawings.

In practical use, the workflow has been designed to generate new images or edit user-provided images conditioned on text or visual input using "in-painting", which is a widespread AI technique that replaces content inside a selected region while preserving surrounding context (Zhang et al. 2023). As such, users load a photo of an interior setting (the *base image*), brush over the area they want to change (the "*mask*"), and then instruct the AI model what to reconstruct in that masked area using text (the "*prompt*") or a *reference image* or both. The rest of the picture stays untouched; in this way, scale, perspective, and context remain unmodified. The tool includes three modes to generate or edit interior imagery:

- *Mode A – Text-driven*. Through text instructions or prompts, the user generates new imagery or brushes a region to edit a given image.
- *Mode B – Reference image-driven*. Through providing an example image or render (for instance, of a particular furniture style or material palette) as a stylistic guide, the user drives the system to edit a region in a given image via masks or generate new imagery.
- *Mode C – Hybrid*. Through combining a textual description with a reference image, users can achieve more precise transfers of style or atmosphere to generate new images or edit regions of given ones.

Across all modes the system can generate several alternatives per run, making it possible to compare and iterate rapidly in a way akin to sketching on tracing paper. Figures 1 and 2 contain diagrammatic presentations of the developed workflows for generating and editing images, respectively. Figure 3 illustrates an example instance of the workflow in the ComfyUI digital environment.

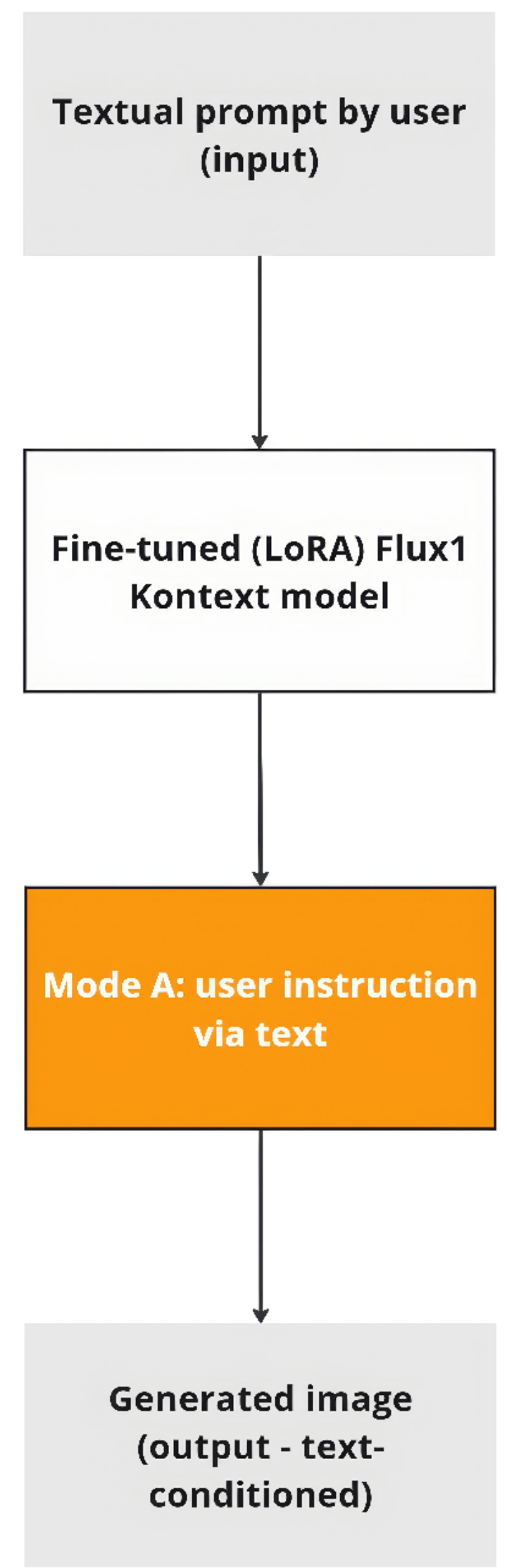

*Figure 1: Our workflow for generating new images using the developed tool*

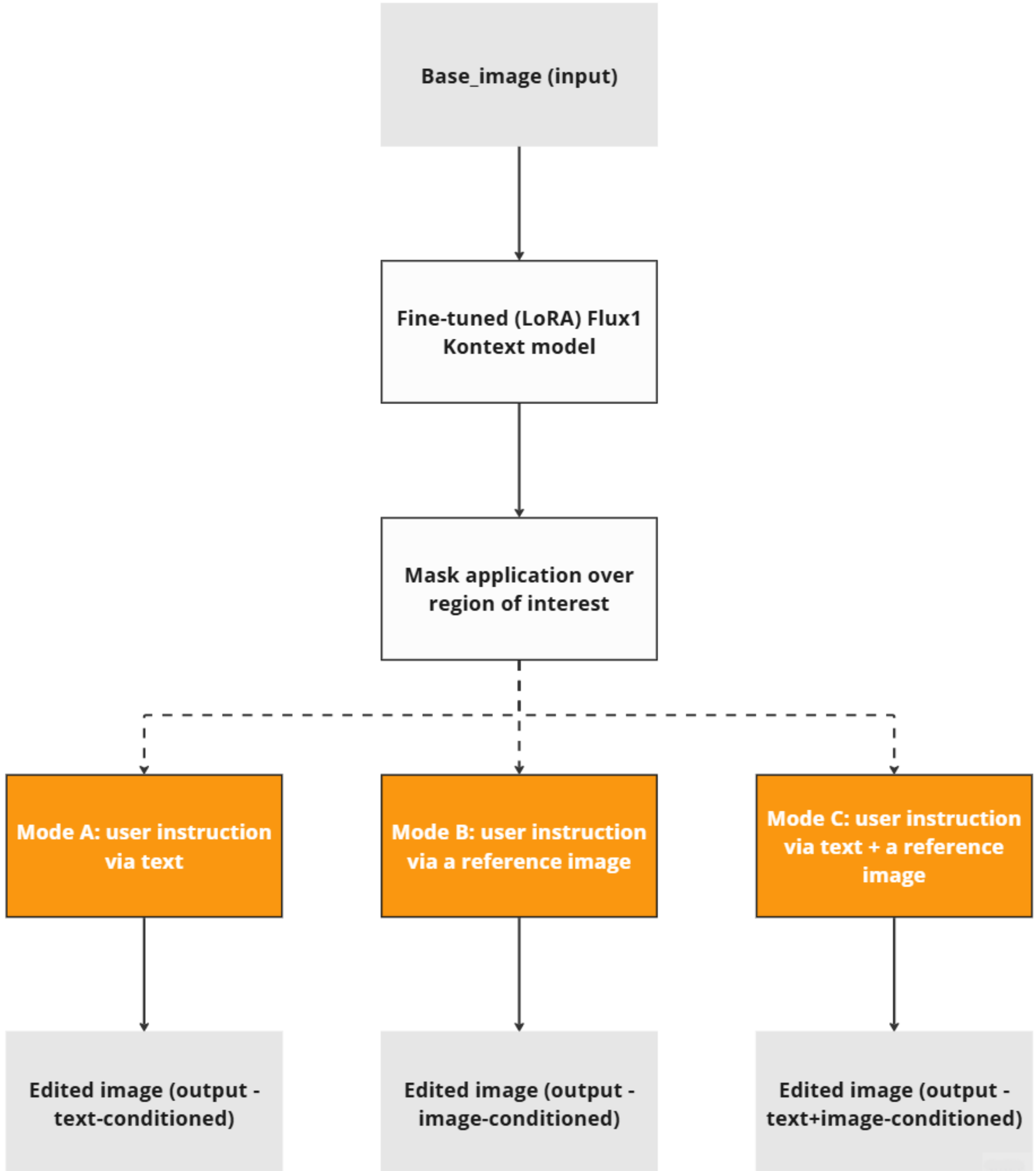

*Figure 2: Our workflow for editing images using the developed tool*

Pedagogically, the tool is aligned with constructivist principles and Universal Design for Learning (UDL). By allowing students to manipulate design ideas directly through visual experimentation and natural language, the system supports learning by doing and provides multiple means of representation and expression. It lowers technical barriers to high-quality visual output, particularly for students who may be less confident with conventional rendering software, while still requiring reflection, judgement and iteration in line with studio pedagogy.

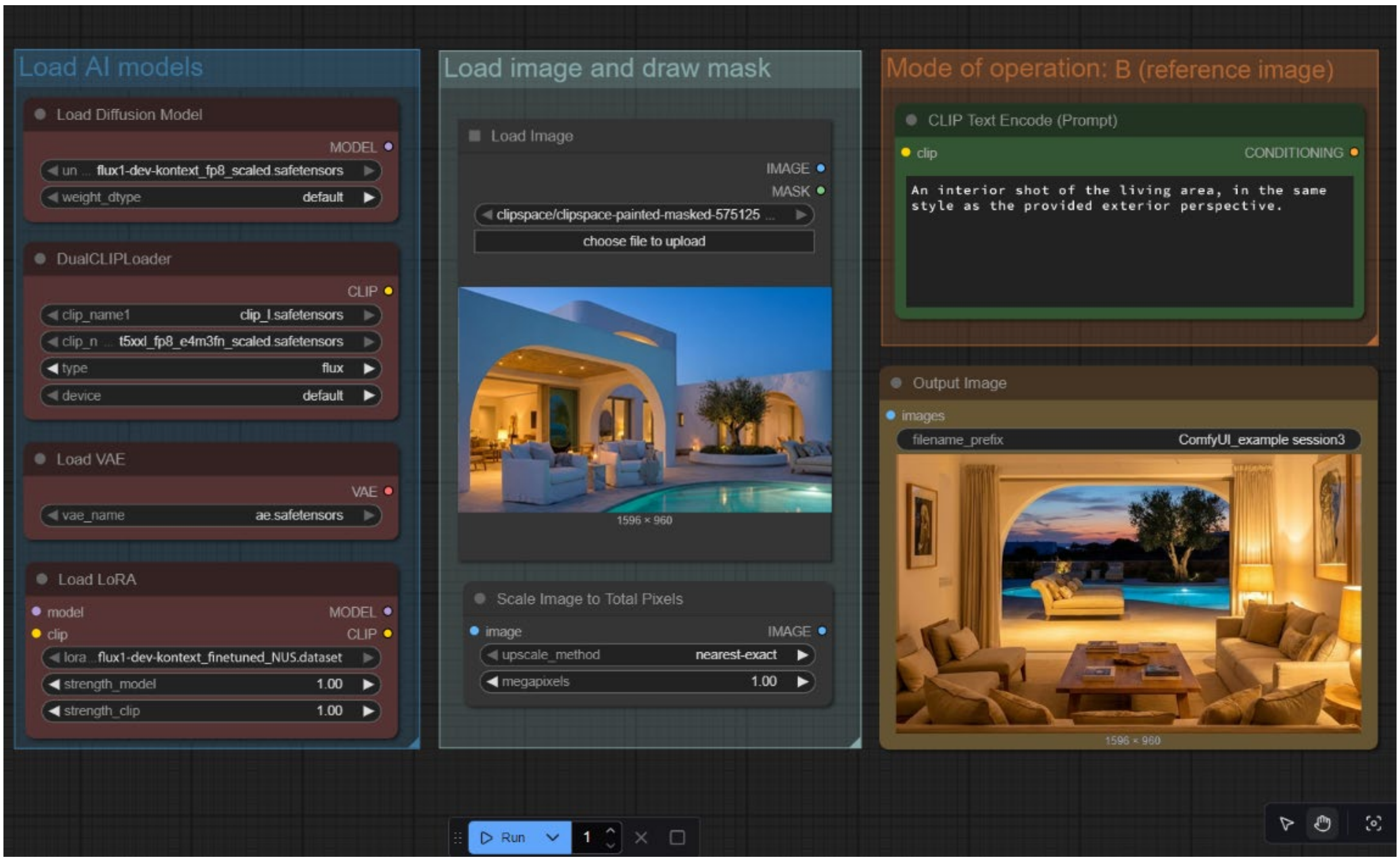

*Figure 3: An example use of the developed tool in a ComfyUI digital environment*

## 2.2. Study design and ethics

The study adopts a focus-group based, mixed-methods design, explicitly structured to evaluate the educational impact of the Gen-AI-tecture tool on creativity, inclusivity and AI-handling skills relevant to employability. Two focus groups, each comprising 5-8 participants, are conducted with Level 3–5 students enrolled on architecture courses at the University of Derby. The cohort thus reflects the actual teaching context in which the tool is deployed.

Sampling is purposive and recruitment is carried out via email invitations and in-class announcements, with the aim of achieving diversity in academic level, international status, declared disability and gender, alongside variation in prior digital design familiarity (for example, experience with CAD or visualisation software). Inclusion criteria were current enrolment on a relevant programme, sufficient English language proficiency to participate in group discussion, and the ability to provide informed consent. There are no exclusion criteria beyond inability to consent.

Each focus-group session is designed to last approximately **90 minutes** and follows a stepwise structure:

a) **Warm-up and consent (5–10 minutes)**
   Participants are welcomed and introduced to the aims of the project as a research-focused evaluation of a teaching innovation, separate from summative assessment. The participant information sheet is reviewed, covering procedures, voluntary participation, the right to withdraw, and audio recording. Written informed consent is collected before any data are generated. A brief ice-breaker invites each student to describe "one memorable digital tool you have used in your design process", establishing rapport and surfacing baseline familiarity with digital media.

b) **Main module – AI image tool exploration (60 minutes)**

- **Showcase (10-15 minutes)**
  The facilitator (i.e., the academic researcher) demonstrates the Gen-AI-tecture workflow in real time, using example prompts (e.g. "Create three interior design variations for a small living room in a typical UK semi-detached house" or "Generate an interior concept for a small building using Brutalist influences") to illustrate the three modes (A-C) and core functions.

- **Interactive prompting – individual work (45-50 minutes)**
  Students then undertake a two-phase design task using the tool, working individually at computer workstations but in a shared space where they can converse informally:

  - **Phase A – Master bedroom (Create)**
    In Phase A, students design a master bedroom from scratch for a fictional house. They are free to experiment with any interior style or atmosphere they wish, using Mode A (Text) to generate new imagery. They are encouraged to iterate, compare multiple AI outputs, and refine their prompts. Towards the end of this phase, they are asked to identify two or three "Phase A final outputs" that best represent their design intent.

  - **Phase B – Living/dining room (Edit)**
    In Phase B, students are provided with CAD-based visualisations and photographs of the living/dining room of the same house. The brief is to adapt this second space so that it is either stylistically coherent with, or deliberately contrasted against, their Phase A bedroom design. Here, students use Mode B (Reference Image) and/or Mode C (Text + Reference) to edit the provided images via in-painting: they can, for example, transfer material palettes, lighting atmospheres or furniture typologies from their Phase A inspirations into the living/dining room, or apply new styling decisions while maintaining spatial consistency. This explicit linkage between Phase A (create) and Phase B (edit) is central to examining how the AI supports both open-ended ideation and context-sensitive refinement.

c) **Reflective discussion and usability (20–25 minutes)**
After the hands-on activity, the group engages in a semi-structured discussion focusing on (i) perceived impacts of the tool on the breadth, depth and originality of their design ideas (creativity); (ii) whether the tool made the design process feel more accessible or inclusive, particularly for those less confident in drawing, language or software; and (iii) how using the tool affected their confidence in managing AI tools more broadly and their sense of preparedness for professional practice (employability). Participants complete the "Usability Evaluation" questionnaire.

d) **Wrap-up and debrief (5 minutes)**
To conclude, participants are invited to offer one or two suggestions for improving the tool. A debrief reiterates the voluntary nature of participation, restates the timeline for withdrawing data, and provides contact details for the researcher, academic supervisor and relevant support services.

Ethically, the study operates under approved institutional protocols (University of Derby Engineering Ethics Committee, reference ETH2425-3722). Participation is strictly voluntary and unrelated to grading decisions. All audio recordings, prompt logs, image outputs and questionnaires are stored on secure, password-protected University servers, accessible only to the research team. Data are pseudonymised using participant codes; names and other direct identifiers are removed from transcripts and reporting. AI-generated images used in dissemination are presented without personal identifiers and are always labelled as "Conceptual Visualisation" to avoid confusion with formal assessed work. The 90-minute session length and structure are designed to minimise fatigue, and the small-group format allows the facilitator to monitor well-being and provide support where needed. Collectively, these measures ensure that the study design is not only aligned with the Study Design document but also proportionate, ethically robust and pedagogically grounded.

## 2.3. Data collection methods and tools

Data was generated using two complementary methods: a *post-session "Usability Evaluation" questionnaire* administered individually to all participants and a *semi-structured, reflexive group discussion* conducted immediately after the hands-on activity. This mixed approach allowed the study to capture both standardised, comparable responses and richer, interactional accounts of students' experiences with the AI tool.

The questionnaire (Table 1) comprised eleven Likert-type items (1 = strongly disagree; 5 = strongly agree), grouped into three subscales aligned with the research objectives – i.e., students' creative skills and outcomes (RO1), inclusivity and equality (RO2), AI-handling skills and employability (RO3) – plus two global items on the overall usability of the session and the tool. These items provided a concise quantitative indication of perceived impact across the three domains of interest.

| Subscale | Item | 1 | 2 | 3 | 4 | 5 |
|---|---|---|---|---|---|---|
| **A – Creative skills and outcomes** | ***A1.*** *The use of the AI tool helped me generate more design ideas than I would normally produce in a similar task.* | O | O | O | O | O |
| | ***A2.*** *The AI tool encouraged me to explore a wider range of styles, materials and atmospheres in my designs.* | O | O | O | O | O |
| | ***A3.*** *The images generated by the AI tool helped me refine and improve my design decisions for the required task.* | O | O | O | O | O |
| **B – Inclusivity and equality** | ***B1.*** *The use of the AI tool accommodated my preferred way of learning and working in design tasks.* | O | O | O | O | O |
| | ***B2.*** *I was able to engage fully with the design task regardless of my previous experience with AI tools.* | O | O | O | O | O |
| | ***B3.*** *The AI tool helped me express my design ideas regardless the level of my sketching or drawing skills.* | O | O | O | O | O |
| **C – AI-handling skills and employability** | ***C1.*** *After this session, I feel more confident using AI tools as part of my design process.* | O | O | O | O | O |
| | ***C2.*** *The skills I practised today will be useful for my future work in architecture or related fields.* | O | O | O | O | O |
| | ***C3.*** *I feel better prepared to discuss and work with AI-supported design workflows with future employers or colleagues.* | O | O | O | O | O |
| **Overall usability** | ***O1.*** *Overall, the structure of today's session (activities, timings, explanations) was easy to follow and participate in.* | O | O | O | O | O |
| | ***O2.*** *Overall, the AI tool was easy and intuitive to use.* | O | O | O | O | O |

*Table 1: A copy of the Usability Evaluation questionnaire administered to participants. Items are rated in an 1-5 Likert scale, with 1=Strongly disagree, 2=Disagree, 3=Neither agree nor disagree, 4=Agree, 5=Strongly agree*

In parallel, a set of open-ended prompts (Table 2) was used to guide the semi-structured, reflexive discussion, inviting participants to elaborate on how the AI workflow influenced their creativity, supported or constrained inclusive participation, and shaped their sense of preparedness for AI-rich professional practice. These discussions were audio-recorded and transcribed verbatim to enable subsequent qualitative analysis.

| Theme | Questions asked to focus-group participants |
|---|---|
| ***Creative skills and outcomes*** | - In what ways, if any, did using the AI image tool influence the number and variety of design ideas you came up with today?<br>- Can you describe a moment when the AI output led you to an idea you would not normally have considered?<br>- How did the AI-generated images affect the way you developed or refined your designs for the two rooms?<br>- Thinking about the final images you selected or liked most, how would you describe their quality and originality compared with work you usually produce without AI?<br>- Do you feel the tool supported your own creative voice, or did it tend to impose a particular "look"? |
| ***Inclusivity and equality*** | - Did the workflow (prompts, iterations, comparisons) make it easier or harder for you to engage with the task?<br>- Were there aspects of the interface or process that felt particularly supportive or, conversely, excluding?<br>- For those with little or no prior AI use: did the session feel approachable, or did you feel at a disadvantage? For those with more experience: did this tool change or confirm your views about AI in design?<br>- Did the tool help you communicate ideas you might struggle to draw or render traditionally? |
| ***AI-handling skills and employability*** | - After today's session, how has your sense of your own AI-handling skills changed, if at all?<br>- Are there particular things you now feel more able to do (e.g. prompt design, critically selecting outputs, iterating with AI)?<br>- Can you imagine specific situations in practice where you might want to use a tool like this?<br>- Are there aspects of professional work (e.g. collaboration with clients, speed of options, visual communication) where you see particular value or risk?<br>- What skills or understandings would you still need to develop to feel fully "employable" in an AI-rich design environment? |

*Table 2: Example questions used in the reflexive, semi-structured discussion with focus groups*

## 3.4. Data analysis methods and tools

Data analysis followed a convergent mixed-methods strategy to examine how the locallyrun GenAI workflow shaped design processes and appraisals across phases. All datasets (questionnaire responses, transcripts, prompt logs, masks and image selections) were keyed by shared identifiers (participant/focus-group IDs, operation mode A/B/C, phase, generation index and room) to support triangulation and auditability.

Qualitative analysis was conducted in *NVivo* (Lumivero, version 15) using reflexive thematic analysis (i.e., familiarisation, coding, theme construction, review, definition and reporting) following Braun and Clarke's (2006) framework, with attention to patterns in how students narrated creativity, inclusivity and AI-handling skills.

Quantitative analysis was carried out in *Python* (Python, version 3.14), using the NumPy and Matplotlib libraries. For each questionnaire subscale we calculated descriptive statistics (means, standard deviations, frequencies and percentage distributions) and visualised them using simple plots.

Finally, to explore convergent validity between questionnaire subscales and qualitative indices derived from the reflexive discussions, we computed Spearman's $\rho$ rank-order correlations with Holm-adjusted p-values within families of related tests. For pairs without ties,

$$\rho = 1 - \frac{6 \sum d_i^2}{n(n^2 - 1)}, \quad (1)$$

where,

$n$ : the number of paired observations being correlated (i.e. the sample size for that specific correlation).
$i$: the index for each observation, running from 1to $n$.
$d_i$: the difference between the two ranks for observation $i$.

# 3. Results

## 3.1. Sample characteristics and interaction with tool

Participants were drawn from Levels 3-5 and allocated to two focus groups (n = 8 per session). The cohort thus reflected a spread of prior studio experience, with a slight predominance of Level 4 students. Around one-third declared a disability, including almost one-fifth of the overall sample reporting a specific learning difficulty, and the ethnic and gender composition was mixed. This configuration aligns with the project's concern to probe inclusivity across diverse learner profiles rather than a narrowly defined subgroup. Table 3 presents the study sample characteristics.

| Characteristic | Category | n | % |
|---|---|---|---|
| **Level of study** | Level 3 | 5 | 31.3 |
| | Level 4 | 7 | 43.8 |
| | Level 5 | 4 | 25.0 |
| **Disability status** | Any declared disability | 5 | 31.3 |
| **(of which)** | Specific learning difficulty* | 3 | 18.8 |
| **Ethnicity** | White | 8 | 50.0 |
| | Asian | 4 | 25.0 |
| | Black | 2 | 12.5 |
| | Mixed / Other | 2 | 12.5 |
| **Gender** | Female | 9 | 56.3 |
| | Male | 6 | 37.5 |
| | Non-binary | 1 | 6.3 |

*Table 3: Sample characteristics (n = 16). (*) e.g. dyslexia, ADHD (self-reported). Percentages may not total 100% due to rounding.*

Log data show that students engaged intensively with the workflow in both focus groups, producing around 80 images per session (approximately ten per student) within the planned 90-minute window. All three operation modes were used substantively, with a modest preference for text-only prompting (Mode A) but continued reliance on reference-image and hybrid modes, indicating that participants combined verbal and visual strategies to steer the AI. Acceptance rates above 60% in both sessions suggest that most outputs were considered valuable enough to retain for critique or further iteration. Table 4 outlines data from students' interactions with the developed tool. Figures 4 and 5 include examples of generated or edited images, which were produced by student participants from both focus groups.

| Variable | Session 1 | Session 2 |
|---|---|---|
| *Total images generated* | 78 | 82 |
| *New images (no base photo)* | 40 | 41 |
| *Edited images (in-painting)* | 38 | 41 |
| *Images generated using Mode A* | 30 | 32 |
| *Images edited using Mode B* | 26 | 24 |
| *Images edited using Mode C* | 22 | 26 |
| *Unique seeds* | 59 | 61 |
| *Acceptance rate* (%)* | 62.8 | 68.3 |
| *Session duration (minutes)* | 84 | 89 |

*Table 4: Interaction with the GenAI tool by focus-group session. (*) Acceptance rate = number of AI outputs that participants chose to keep, annotate or build on / total images generated in that session.*

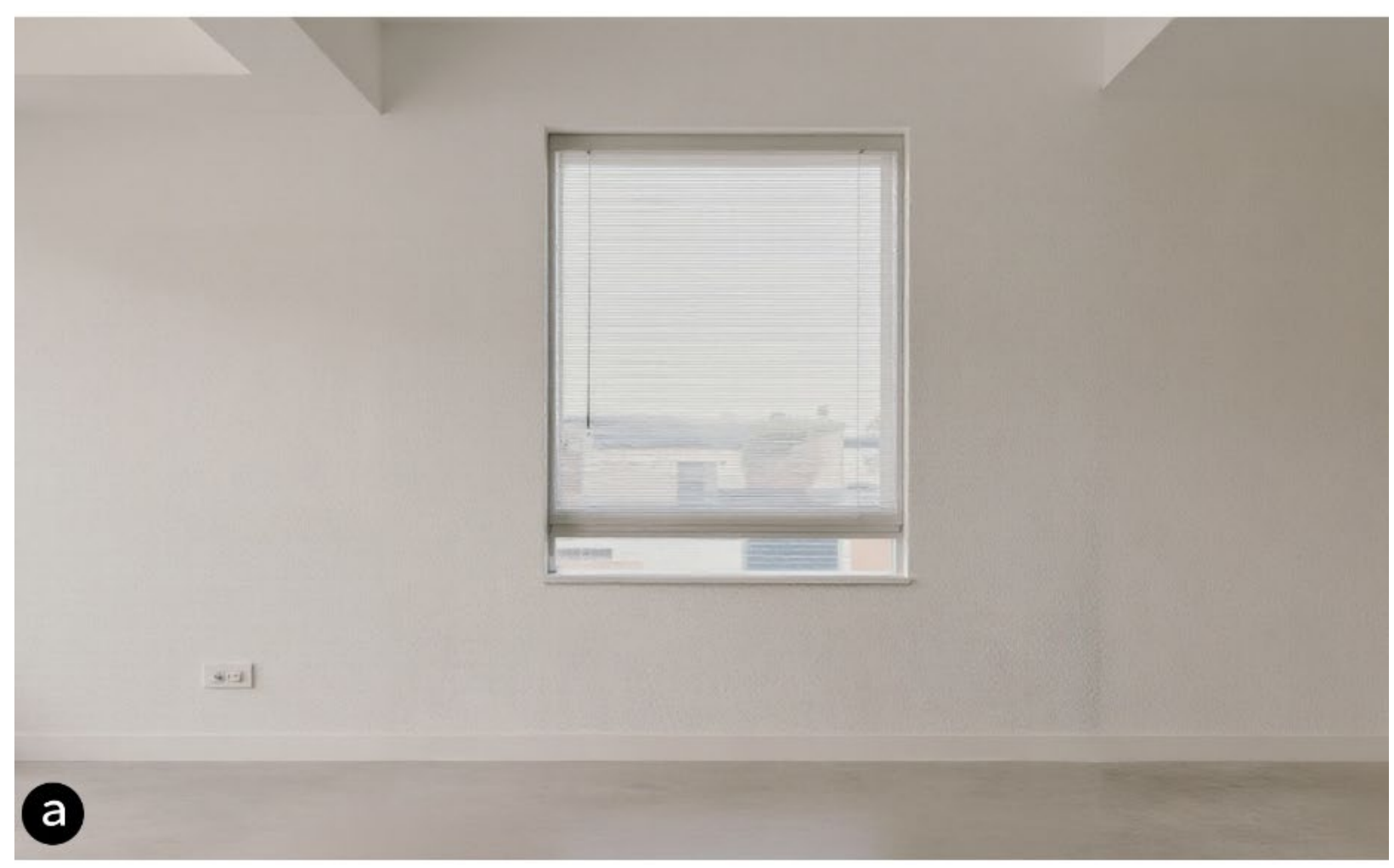

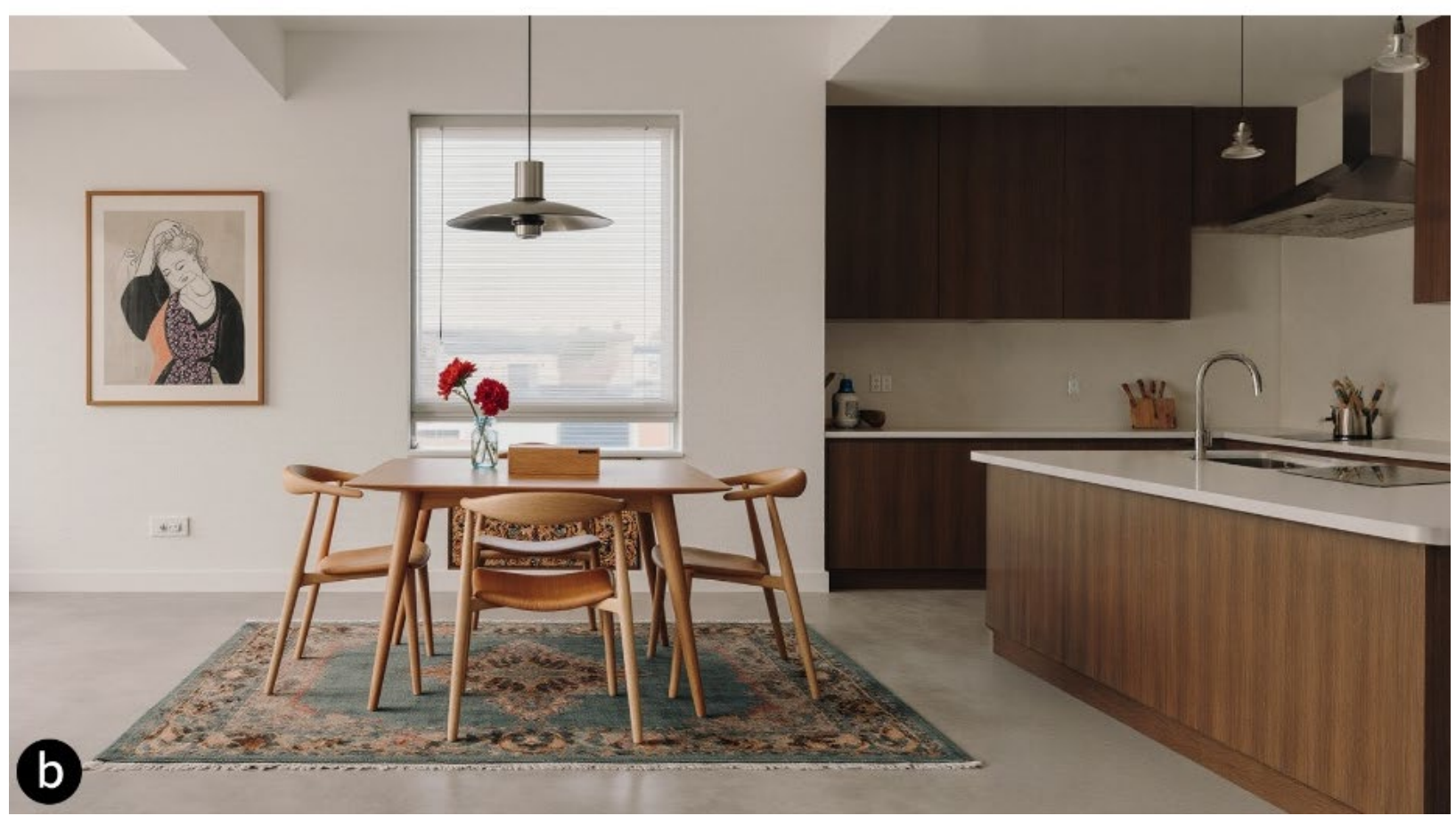

*Figure 4: An example output from focus groups. Students were given an image of an empty dining room (a), which they then edited using the developed tool (b).*

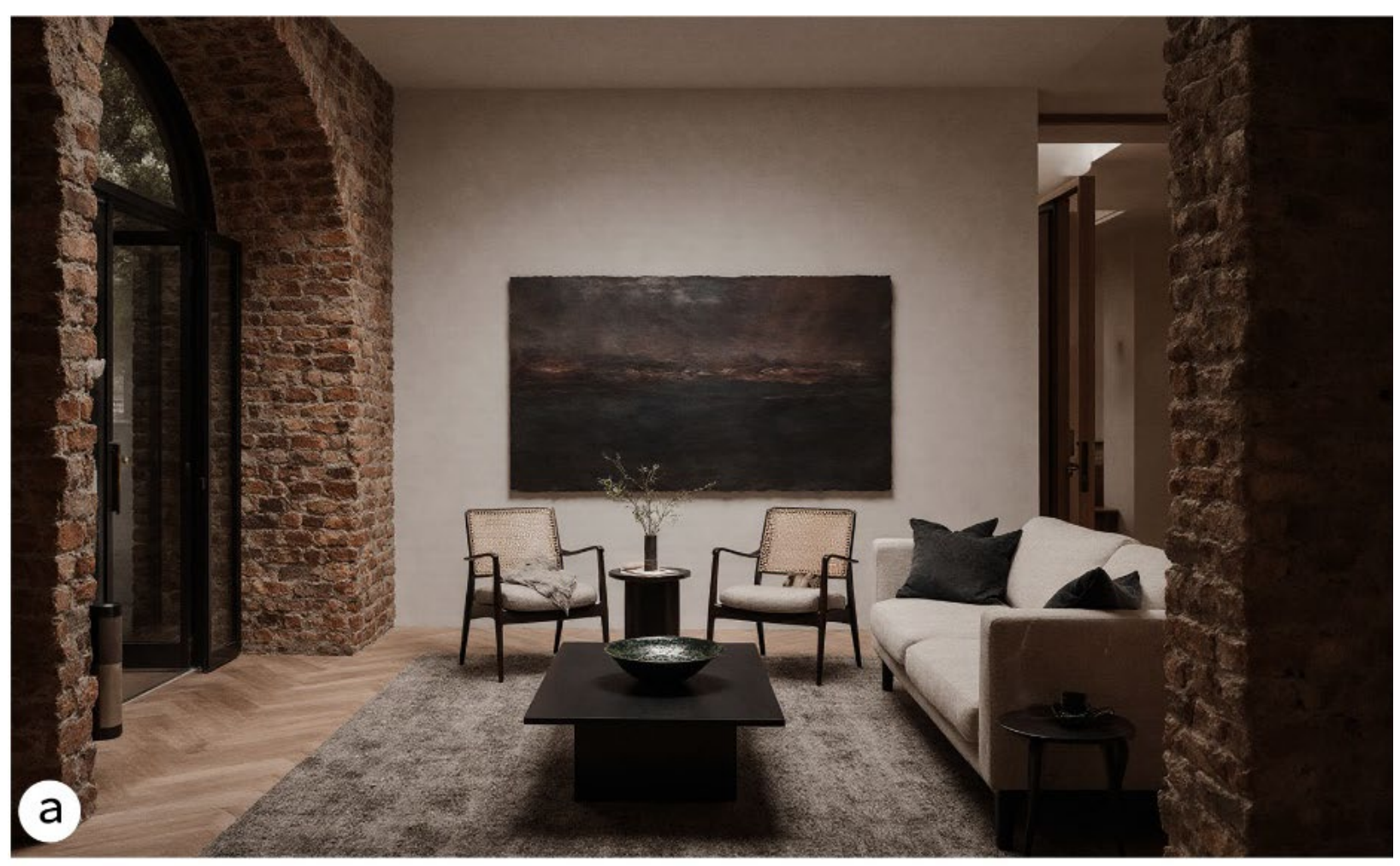

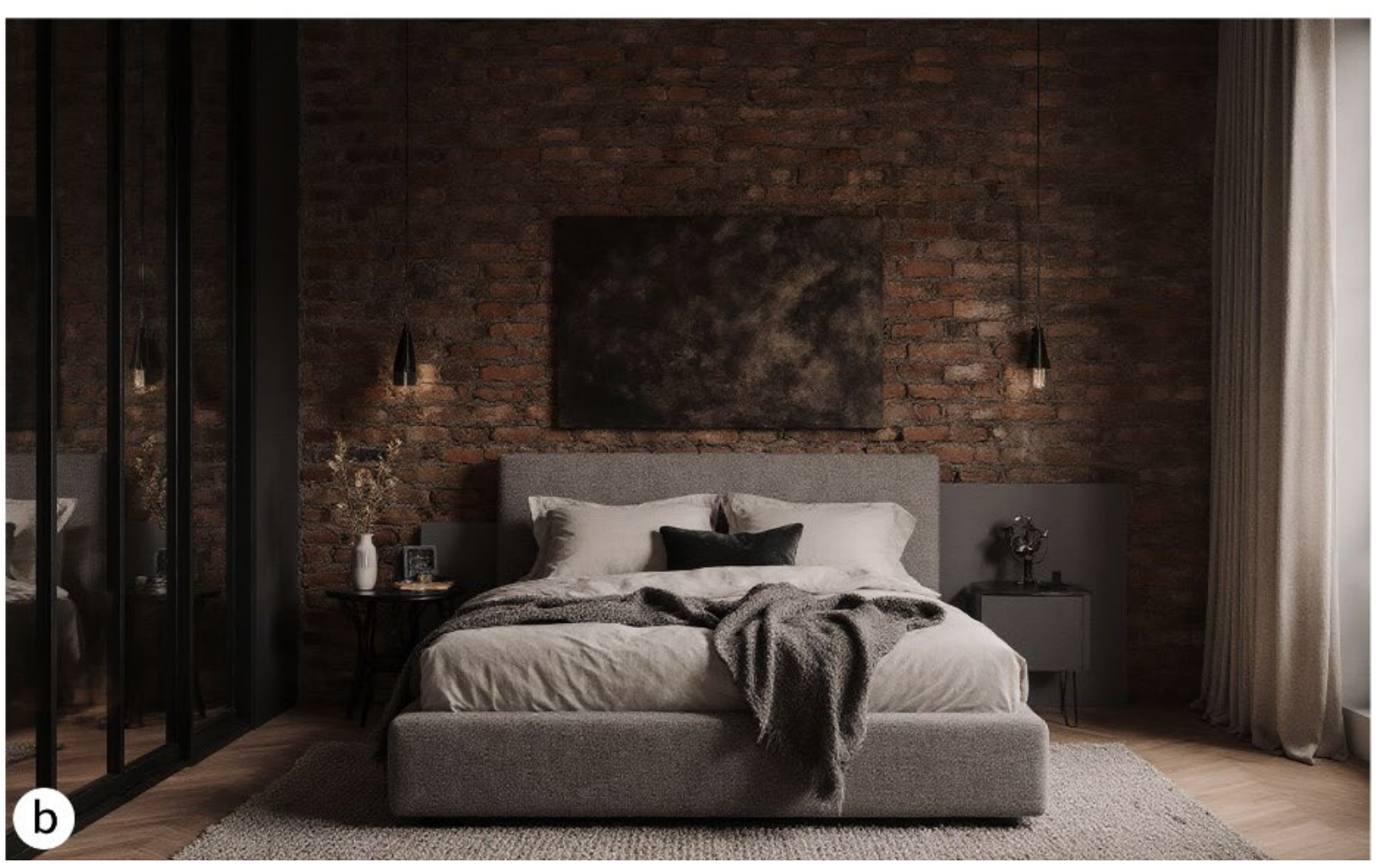

*Figure 5: An example output from focus groups. Students generated an image of a living room (a), and then used it as a reference to transfer the interior style to a bedroom area (b).*

## 3.2. Objective-related findings

Questionnaire analysis (Figure 6) across both sessions indicated consistently high ratings on creativity-related items (A1-A3), with particularly strong endorsement for the tool as a stimulus for new ideas (A1 = 4.5) and experimentation (A2 = 4.2), and a slightly lower, but still positive, score for depth of exploration (A3 = 3.7). Inclusivity-oriented items (B1-B3) cluster around the upper-mid range (3.8-4.1), suggesting that students generally experienced the workflow as supportive of participation, confidence and access, rather than exacerbating existing inequalities. AI-handling and process items (C1-C3) show a more differentiated pattern: C1 and C2 remain positive (3.7 and 3.5), whereas C3 (concerning confidence in transferring these skills beyond the studio) is notably lower (2.6), pointing to residual uncertainty about external applicability. Overall, outcome items (O1-O2) sit in the high range (3.9 and 4.2), implying that students perceived the bespoke GenAI workflow as beneficial for their learning and as contributing, at least in part, to their future employability.

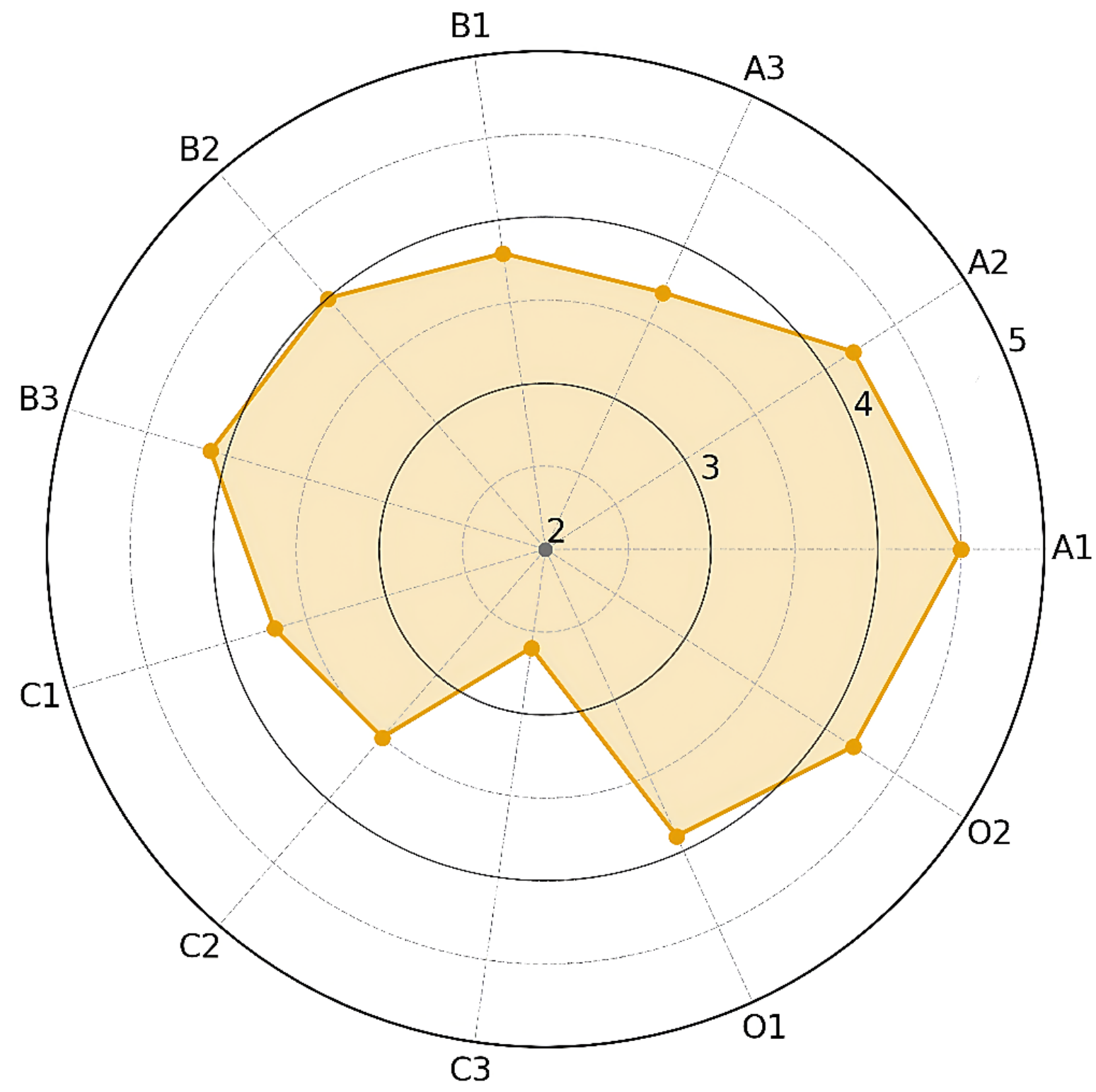

*Figure 6: Questionnaire item-level mean scores across both focus-group sessions*

### 3.2.1. Creative skills and outcomes

The creative-skills questionnaire subscale (Figure 6; A1-A3) showed some of the highest scores, indicating that students generally felt the GenAI workflow enhanced idea generation and design development. At participant level, the mean A-subscale score correlated strongly with qualitative indices for "more ideas than usual" ($\rho = .76$, $p < .01$) and "trying unfamiliar styles/materials" ($\rho = .71$, $p < .01$), evidencing convergent validity between questionnaire ratings and reflexive accounts. Students scoring highly on A1 frequently described expanded ideation: "normally I would stop after two or three options, but with the AI I felt I could quickly push five or six quite different room layouts before choosing one" (Participant 03/Group 1). By contrast, a lower-scoring student remarked that "It gave me some variations, but they were quite similar to what I already had in mind" (P06/G2).

Higher A2 scores aligned with richer descriptions of stylistic experimentation; participants reported using the tool to "test atmospheres I do not usually draw, like very dark, moody lighting or much richer textures" (P06/G1) or to "swap between quite minimal and very layered interiors without re-modelling everything" (P05/G2), with a very strong correlation between A2 and the index "experimenting beyond usual repertoire" ($\rho = .79$, $p < .01$). A3, concerning refinement of design decisions, was more moderate (3.7) and thematically nuanced. Some students noted that "the AI helped me spot combinations I had not thought of, but I still had to decide what really works for the brief" (P02/G1), while others felt that "Sometimes it refined things, other times it distracted me with details that looked cool but did not fit the concept" (P08/G2). Still, A3 correlated strongly with references to "using AI images to justify or adjust final choices" ($\rho = .64$, $p < .01$). Overall, the convergent quantitative and qualitative evidence suggests that the workflow most robustly expanded students' creative search space, while supporting, though less uniformly, the refinement of final design decisions.

### 3.2.2. Inclusivity and equality

Questionnaire mean scores on the inclusivity subscale (Figure 6; B1-B3) were consistently high, indicating broadly positive perceptions of accessibility and equitable participation. The mean B-subscale score correlated strongly with qualitative indices for feeling the session was "approachable/welcoming" ($\rho = .74$, $p < .01$) and for "keeping up regardless of prior AI use" ($\rho = .81$, $p < .001$), evidencing clear convergent validity between questionnaire ratings and reflexive narratives.

For B1, higher scores aligned with accounts of the workflow fitting students' preferred ways of working. As one participant observed, "*I liked that I could talk through ideas first, then use the prompts almost like annotations on my sketches rather than starting from a blank screen*" (P02/G1). Another contrasted the format with conventional crits: "*Usually I feel a bit on the spot in crits, but here I could try things quietly, then only show the versions I was happy with*" (P04/G2).

B2 and B3, focused on equitable engagement and expression regardless of prior AI or drawing skills, also showed strong correlations with qualitative indices ($\rho = .78$ and $\rho = .72$, respectively, $p < .01$). Students with limited AI experience frequently reported not feeling disadvantaged: "*I came in worried that people who already use Midjourney would dominate, but the way the steps were broken down meant I did not feel behind*" (P01/G1). Those less confident in drawing similarly described the tool as levelling: "*Normally my rough sketches do not look like what is in my head, but here the AI helped me show the idea without needing perfect linework*" (P02/G2), and "*It felt fairer because we were all editing photos rather than comparing who can render best*" (P07/G1). Together, these convergent quantitative and qualitative signals suggest that the bespoke GenAI workflow operated as a modest equalising mechanism, cultivating a culture of inclusion in the classroom.

### 3.2.3. AI-handling skills and employability

Scores on the AI-handling and employability subscale (Figure 6; C1-C3) were more moderate than in other domains (C1 = 3.7; C2 = 3.5; C3 = 2.6), indicating increased confidence in using the tool within the session but more hesitation about longer-term transfer and employment. The mean C-subscale score correlated strongly with a qualitative index on "clearer step-by-step understanding of how to work with AI" ($\rho = .71$, $p < .01$), supporting convergent validity between ratings and reported procedural learning.

Higher C1 scores aligned with accounts of strengthened workflow literacy. As one participant reflected, "*Before today I just typed random prompts into whatever AI was available; now I feel I understand how to structure the prompt and tweak things instead of starting over each time*" (P04/G1). Another observed that "*Seeing the mask, prompt and seed together helped me think of the AI as another tool in the workflow, not a magic black box*" (P05/G2), comments that were frequently co-coded with a "procedural confidence" index strongly associated with C1 ($\rho = .68$, $p < .01$).

C2 and C3, focused on future usefulness and employability, were more cautious but still positively associated with relevant qualitative indices. Higher C2 scores correlated strongly with references to concrete professional scenarios ($\rho = .72$, $p < .01$), such as "*quick visuals for client meetings when you need to test alternative finishes on the spot*" (P06/G2) or "*mocking up options for retrofit projects without doing full 3D models straight away*" (P08/G1). The lower mean on C3 matched more tentative employability narratives, despite a strong correlation with an index on "awareness of employer expectations around AI" ($\rho = .63$, $p < .05$). As one student noted, "*I can see this kind of tool becoming standard in offices, but I am not yet sure I could explain the technical side confidently in an interview*" (P03/G1), while another added, "*It makes me feel slightly more prepared, but I also realise how much more I would need to learn about data, ethics and settings before using it with real clients*" (P07/G2).

Overall, the convergent quantitative and qualitative evidence suggests that the session positively contributed to immediate AI-handling skills and prompted reflection on AI-related employability, while also making visible perceived gaps that would require more sustained, curriculum-level provision.

# 4. Discussion

Our findings suggest that the bespoke, locally run GenAI workflow functions as a constructionist "*microworld*" for architectural ideation rather than as a mere image generator. Strong correlations between the creative-process subscales and the qualitative indices of divergent exploration, iterative refinement and risk-taking in the reflexive discussions indicate that students who rated the tool as enhancing their creativity were also those who, in practice, used it to branch, recombine and challenge initial design ideas. This resonates with Papert's notion of digital artefacts as "*objects-to-think-with*" and subsequent work on AI-rich microworlds, where learning is evidenced by cycles of making, inspecting and remaking rather than by converging on a single "correct" outcome (Papert, 1980; Kahn and Winters, 2021). In line with art-education studies where GenAI images become stimuli for critical and experiential learning (Pavlik and Pavlik, 2024; Bian et al., 2025), participants in this study consistently described AI-edited images as provocations that "opened up" unforeseen options, especially when switching between the different masking modes. The present findings extend this line of work into architectural pedagogy by showing that a spatially constrained, context-preserving in-painting workflow can sustain constructionist cycles of experimentation at the scale of streets, interiors and public spaces, while remaining legible as a design process rather than as opaque automation. In this sense, the tool aligns with constructivist conceptions of AI as a cognitive artefact that supports active knowledge construction and design reasoning (Gibson et al., 2023; Guo, Yi and Liu, 2024; Gibson and Ifenthaler, 2024).

At the same time, the data point towards a distinctly socio-constructivist reading of students' creative engagement with the system. Reflexive accounts indicated that the AI often acted as a "*more knowledgeable other*" that suggested plausible yet negotiable alternatives; prompting students to articulate constraints, precedents and values more explicitly, particularly among those who initially reported lower confidence in drawing or visualisation. This pattern, again supported by strong positive correlations between perceived creative support and coded episodes of explanation, justification and self-correction, echoes recent work on GenAI as a scaffold within Vygotskian frameworks, extending learners' zones of proximal development when its outputs are interrogated rather than accepted uncritically (Tran et al., 2025; Sánchez Muñoz et al., 2025). The workflow thereby operates as a cognitive mediator that elicits metacognitive talk about design moves and trade-offs, in a way that parallels socratic AI tutors in text-based domains (Degen and Asanov, 2025) and inquiry-oriented uses of GenAI in other disciplines (Moundridou, Matzakos and Doukakis, 2024; Kotsis, 2024). Crucially, the locally run, human-in-the-loop configuration observed here appears to preserve learner agency; students framed the AI as a fast, suggestive collaborator rather than a judge, and instances where AI outputs risked narrowing exploration were typically identified and problematised in group discussion. In this respect, the study contributes an empirically grounded example of how a domain-specific GenAI tool can operationalise human-centred, constructivist principles in studio-based architectural education (Anastasiades et al., 2024; Ouyang and Jiao, 2021; Owen, 2025), demonstrating that quantitative gains in perceived creativity are intertwined with richer dialogic and reflective practices around design.

Results of this study suggest that the developed tool acted as a concrete instantiation of Universal Design for Learning (UDL) principles within an architectural context. Throughout the study, participants expressed increased confidence to share work, perceived "levelling of the playing field", and reduced anxiety in crit-type situations. This is closely aligned with UDL's emphasis on multiple means of representation, engagement and action/expression (Meyer, Rose and Gordon, 2014) and with recent arguments that AI can operationalise UDL by diversifying media and pacing while supporting learner autonomy (Saborío-Taylor and Rojas-Ramírez, 2024; Hyatt and Owen, 2024). In our case, the combination of masked editing, text prompts and reference imagery provided alternative routes into spatial thinking for students who reported difficulties with freehand drawing, technical English or traditional orthographic representation, echoing accounts of AI as an assistive UDL-aligned technology for disabled and neurodivergent learners (Ayala, 2024; Pack, 2024). Importantly, students did not describe the workflow as a bolt-on accommodation but as part of the "normal" design process, which resonates with UDL's proactive, whole-cohort framing and with work on digital technologies as enablers of inclusive design when deliberately aligned with UDL checkpoints rather than retrofitted after the fact (Veytia Bucheli et al., 2024; Meyer, Rose and Gordon, 2014). In this sense, the study extends largely text-focused AI-UDL research into the visual and spatial domain, demonstrating that a locally run, fine-tuned image workflow can support inclusive participation in design ideation by lowering representational and affective barriers while maintaining students' authorship over design intent.

In contrast to strands of the inclusive-education literature that warn AI may exacerbate existing inequities when layered onto already stratified classrooms (Melo-López et al., 2025; Gupta and Kaul, 2024; Pagliara, 2024), our data suggest that a carefully scaffolded, locally controlled GenAI workflow, embedded within a UDL-informed session design, can enable students to "keep up" irrespective of prior AI or drawing experience. This finding complements work on AI-based education tools as enablers of inclusive education (Šumak et al., 2024; Temirov et al., 2025) by providing fine-grained evidence from an architectural studio that students who entered the session worried about AI or their visual skills did not report feeling disadvantaged and, in many cases, explicitly characterised the format as more equitable because it focused on shared photo-editing tasks rather than on comparative rendering ability. In doing so, the project offers a practice-based illustration of how inclusion-oriented AI design - attentive to access, representation and learner citizenship (Dieterle, Dede and Walker, 2022; Roshanaei, 2024) - can translate into *experienced* inclusion at the level of concrete design tasks, thereby enriching largely conceptual policy debates on AI, equity and participation in higher education (Varsik and Vosberg, 2024).

Our findings indicate that students' developing AI-handling skills were closely bound up with their capacity to build, navigate and critically manage learning networks, in a manner that is well captured by connectivist theory. Specifically, it was reported that students were not simply "using a feature" but actively orchestrating connections between the GenAI workflow, their own design ideas, peer feedback and external precedents. This resonates with Siemens' (2005) and Downes' (2022) view of learning as the ongoing configuration of networks linking human and non-human nodes, where knowledge resides partly in "appliances" and learning involves pattern recognition and decision-making across those networks. Students' reflexive descriptions of moving iteratively between sketches, textual prompts, masked edits and ensuing AI variants align with the kind of connectivist "meta-skills" highlighted by Mukhlis et al. (2024), filtering, evaluating and curating information flows in the digital age. In this sense, the bespoke, locally run GenAI workflow functioned as an interaction agent within students' personal learning environments rather than an opaque tutor, foregrounding what Downes (2022) later calls *personal* rather than merely *personalised* learning, and echoing reports from other higher education contexts in which GenAI models are experienced as one node in a wider ecology of lectures, labs, peers and online sources (Gottipati, Shim and Shankararaman, 2023; De Silva et al., 2025).

Quantitative and qualitative signals around perceived employability likely interpret AI-handling skills as precisely the kind of connectivist competencies that contemporary practice in architecture and the built environment is likely to demand. The developed workflow provided students with concrete strategies for integrating AI techniques into broader professional processes, rather than treating GenAI as a stand-alone shortcut. For example, using the masking and in-painting modes to probe design opportunity spaces, exploiting its rapid iteration cycles to test variations, and drawing on its traceable outputs to cross-check AI suggestions against professional standards and anticipated team input. This pattern parallels conceptual work in other domains that frames generative AI as an "interaction facilitator" and "interaction agent" in networked learning and professional ecosystems (Liang and Bai, 2025; Yu, 2025), and aligns with reviews arguing that employability in AI-saturated fields increasingly depends on the ability to orchestrate human-AI networks rather than to master isolated tools (Topham et al., 2025). Within a connectivist lens, the study thus contributes fine-grained evidence that a domain-specific GenAI workflow, embedded in a studio setting, can foster students' capacity to construct and manage meaningful connections between AI outputs, disciplinary knowledge and collaborative practices. Rather than undermining expertise, the locally controlled, human-in-the-loop configuration appears to have enabled students to rehearse the connective, evaluative and ethical judgements that underpin both connectivist learning and future-facing professional practice.

This study has several limitations that should be acknowledged and that point towards directions for future work. First, the focus-group-based, mixed-methods design was implemented with a relatively small, purposively sampled cohort of Level 3-5 students from a single UK institution, which constrains statistical power and limits the transferability of the findings to other programmes, disciplines and institutional contexts. Second, the evaluation captured only a single 90-minute exposure to the Gen-AI-tecture workflow and relied primarily on self-report questionnaires and reflexive group discussion. As such, the durability of perceived gains in creativity, inclusivity and AI-handling skills, and their relationship to objective performance indicators (e.g. assessed work, studio grades, portfolio quality) remain uncertain. Third, the bespoke workflow was fine-tuned for residential interior scenarios and deployed within a carefully scaffolded session, which may not straightforwardly generalise to other design typologies, levels of complexity or less supported studio environments.

Future research should therefore extend this work through larger, multi-cohort and multi-institutional studies; include comparison conditions using generic GenAI tools and non-AI teaching formats; and embed the workflow longitudinally across modules to examine its impact on assessment outcomes, portfolio development, progression and retention. Longitudinal designs that follow students into placement or early professional practice would allow closer examination of how AI-handling skills rehearsed in the studio translate into employability. Additionally, co-designing further iterations of both the tool and the learning activities with disabled, neurodivergent and otherwise under-represented students, as well as with industry partners, and undertaking more fine-grained analysis of prompt logs and image trajectories, could deepen understanding of inclusivity, support iterative refinement of the workflow, and yield richer insights into evolving design strategies over time.